\documentclass[final,5p,times,twocolumn]{elsarticle}

\usepackage{graphicx}
\usepackage{amsmath,amssymb,amsfonts}
\pdfmapfile{+txfonts.map} 

\journal{Current Opinion in Solid State $\&$ Materials Science}

\begin{document}

\begin{frontmatter}

\title{Point Contact Spectroscopy of Fe Pnictides $\&$ Chalcogenides In The Normal State}

\author[A]{Hamood Z. Arham}
\author[A]{Laura H. Greene} 
\address[A]{Department of Physics and the Frederick Seitz Material Research Laboratory, University of Illinois at Urbana-Champaign, Urbana, Illinois 61801, USA}

\begin{abstract}

We review the current status of point contact spectroscopy on the iron based superconductors, focusing on their normal state. Point contact spectroscopy is generally used to study superconductors via Andreev reflection, but in recent years it has also proved to be a useful bulk probe of strongly correlated electron systems. Point contact spectroscopy picks up a conductance enhancement in the normal state, above the structural phase transition, of certain iron based compounds. These include Co doped $\rm{BaFe_2As_2}$, $\rm{SrFe_2As_2}$, $\rm{Fe_{1+y}Te}$ and F doped $\rm{SmFeAsO}$ and $\rm{LaFeAsO}$. Two materials which do not show this conductance enhancement are $\rm{CaFe_2As_2}$ and K doped $\rm{BaFe_2As_2}$. This conductance enhancement is thought to be tied to orbital fluctuations. Orbital fluctuations in the normal state of these compounds increases the single particle density of states at the Fermi level, indicating that PCS is sensitive to this excess density of states. The enhancement is only observed at those temperatures and dopings where an in-plane resisitve anisotropy in the detwinned compounds is known to occur. Thus point contact spectroscopy provides strong indications of electronic nematicity in such materials. We also present diagnostics on how to judge if a junction is impacted by joule heating or not. We conclude with the outstanding challenges in the field and the new experiments that need to be carried out. 

\end{abstract}

\end{frontmatter}

\section{Introduction}

Novel and high temperature superconductivity often competes with and appears in close proximity to strongly correlated electron phases. Doniach phase diagrams \cite{Doniach} are helpful in visualizing the competing phenomena in such a compound. At one end of the phase diagram, the compound is in a correlated non-Fermi liquid like state. This may be a Mott insulating, orbital ordering, spin density wave, charge density wave, heavy fermion or pseudo-gap phases. The tuning parameter is generally electron/hole concentration by doping, pressure, or strain. As the tuning parameter is varied, the strongly correlated phase is suppressed and at low temperatures superconductivity emerges. Further variation in the tuning parameter reduces the electron correlations even further giving rise to a strange metallic phase before eventually crossing over into a Fermi liquid like state. For some compounds these phases are mutually exclusive while for others superconductivity may coexist with the preceding strongly correlated phase. Figure 1 is a simple picture portraying such a phase diagram and is equally applicable to copper based superconductors, heavy fermion superconductors, and iron based superconductors. An understanding of the correlated state would most certainly shed light on the mechanisms of novel superconductivity, and therefore may help in our efforts to design new high-temperature superconductors.  

The low temperature ground state of the parent compounds of the iron based superconductors is an antiferromagnetic spin density wave with an orthorhombic or monoclinic crystal lattice. Above the magnetostructural transition ($T_N/T_S$), they are paramagnets with tetragonal crystal lattice \cite{Cruz,Rotter}. It is not clear if this transition is driven by magnetic fluctuations \cite{Fang,Fernandes} or orbital ordering \cite{Chen,Lv,Lv1,Lee}. The suppression of this antiferromagnetic state by various means causes superconductivity to emerge \cite{Ren}. In certain families of the iron based compounds superconductivity and antiferromagnetism coexist. There is evidence that the quantum critical fluctuations associated with the magnetostructural transition are nematic in character and extend in to the normal state of these compounds \cite{Chu, Harriger}. Point contact spectroscopy (PCS) has been found to be sensitive to these nematic electron phases. In this review we summarize the results of PCS measurements in the normal state of the iron based superconductors. Section 2 gives a brief introduction to the technique of point contact spectroscopy, Section 3 presents and discusses the experimental results obtained, along with the theoretical efforts made to interpret them, Section 4 talks about non-ideal point contact junctions, and we conclude in Section 5 by enumerating the insights PCS has provided to the field, additional experiments that need to be performed and the outstanding challenges with regards to the interpretation of PCS data.    

\begin{figure}[hpbt]
		\includegraphics[scale=0.35]{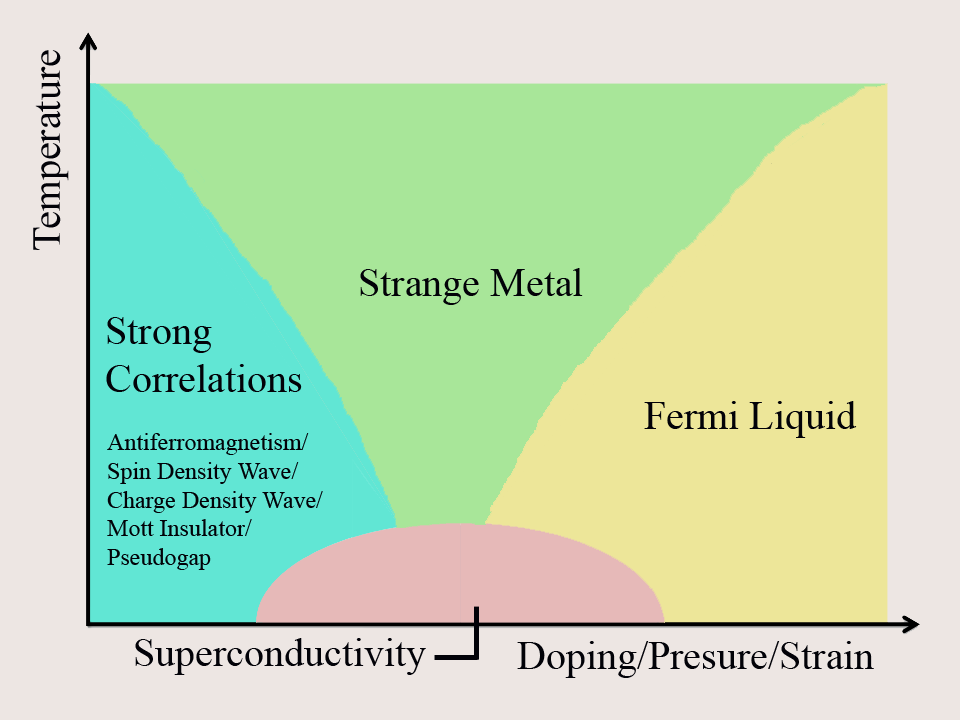}
	\caption{A simplified representative phase diagram for high temperature superconductors. The parent compounds are in a strongly correlated state. When this state is suppressed by various means (doping/pressure/strain) superconductivity emerges. Upon further change in the tuning parameter, the compounds eventually crossover into a Fermi liquid like state. To understand the microscopic mechanism driving the superconducting transition, it is essential to consider its relationship with the correlated state that precedes it.} 
	\label{fig:311}
\end{figure}  

\section{Point Contact Spectroscopy}

Planar tunneling into superconducting Pb gives nonlinearities in the conductance corresponding to the Eliashberg function, $\alpha^2F(\omega$), as was first demonstrated by McMillan and Rowell in 1965 \cite{Rowell,Rowell1}. If the Pb is driven normal (by applying magnetic field or raising temperature), conductance is constant as per Harrison's theorem \cite{Harrison}. (Harrison's theorem shows that when planar tunneling into a Fermi liquid at low bias, the Fermi velocity divides out the density of states.) In 1974, however, Yanson found nonlinearities corresponding to $\alpha^2F(\omega$) in a shorted Pb planar tunnel junction where the Pb had been driven normal by applied magnetic field \cite{Yanson}. Yanson and co-workers showed that their observed nonlinearities arose from the nano-shorts, or ``point contacts'' through the tunnel barrier. The measurement, therefore, was not tunneling, but quasiparticle scattering; hence PCS is also called quasiparticle scattering spectroscopy (QPS).

It is easy to understand what is detected in PCS or QPS through a simple, non-quantum mechanical picture. There are three size regimes of a metallic junction: ballistic, also called the Sharvin limit, where the junction is smaller than the electron scattering length; diffusive, where the junction size is between the elastic and inelastic scattering lengths; and thermal, where the junction is larger than the electron mean free path. There are many good reviews describing these regimes \cite{Park, Daghero, Daghero1}, so we simply point out the basics here. In the thermal regime, the junction acts like a simple resistor and spectroscopic information cannot be derived. In the ballistic regime, electrons are injected a scattering length into the bulk of the sample, and the Eliashberg function is detected when the electron is inelastically backscattered into the orifice: scattering is detected as a slight decrease in the conductance due to the backflow of electrons. This is a small effect and is detected as dips in the second harmonic $d^2I/dV^2$. In the diffusive regime, there is some spectroscopic information, but depending on how close or far the junction is from the ballistic/thermal limit, the spectra can exhibit a range of smearing. Section 4 of this review compares and contrasts thermal and non-thermal limit data. We also point out an important diagnostic: observing signatures apparent in the resistivity (such as phase transitions) can indicate that the junction is in the thermal limit.

The theory for PCS as a spectroscopic technique for quasiparticle scattering off excitation modes (e.g., phonons, magnons) has been well developed \cite{Duif, Naidyuk}. For single-band s-wave superconductors, the theory is also well established. The seminal work of Blonder, Tinkham, and Klapwijk (BTK) \cite{Blonder} shows how to map out the superconducting density of states from the tunneling to the metallic (Andreev reflection) \cite{Andreev} limits. We stress that the data obtained in the tunneling and Andreev limits look completely different, but with the correct BTK analysis, one can obtain the same spectroscopic results: the gap, phonons, and in the case of extending the BTK theory, the order-parameter symmetry \cite{Tanaka,Park1}.

However, there is a lack of theoretical work explaining PCS spectra on novel, strongly correlated materials with nontrivial electron matter. There is no Fermi golden rule type of theory indicating that an increased density of states would yield a larger junction conductance. Nowack and Klug \cite{Nowack} did show how it is possible to explain energy-dependent density of states (DOS) with standard first-order ballistic point-contact theory, but as they state, ``a theory treating electronic DOS and scattering as interconnected would be preferable.''

Recent work has shown that the PCS technique detects density of states arising from strong electron correlations. For heavy fermion compounds, the onset of the Kondo lattice appears as a Fano line shape in the PCS spectra \cite{Park2}. PCS is also sensitive to the hybridization gap and Fano resonance in the heavy fermion $\rm{URu_2Si_2}$ \cite{Park3}. 

In the iron based superconductors, Arham et al. \cite{Arham} have shown that an increase in the zero-bias conductance can be associated with an increase in the single-particle density of states arising from the onset of orbital ordering fluctuations. This not only shows that these materials do indeed exhibit such ``electron matter'' but also that PCS is a powerful bulk probe of such states.

PCS is carried out in two different configurations: the needle anvil method and the soft PCS method \cite{Daghero1,Naidyuk}. In the needle-anvil method, an electrochemically sharpened or mechanically cut metallic tip is brought into gentle contact with the sample to serve as the counter electrode. Gold, silver or platinum wires are normally used as counter electrodes because of their inertness. In the soft PCS method, the counter electrode is mechanically stuck to the sample using conducting silver paint or epoxy. Sometimes, an insulating layer of aluminum oxide is deposited on the sample before affixing the counter electrode. Parallel, nanoscale channels are introduced for ballistic current transport by fritting \cite{Holm} across the oxide layer. The needle-anvil method gives more control over the junction resistance while soft PCS has the advantage of being more stable with temperature change. 

\section{Experimental Results}

Point contact spectroscopy has been used extensively to study the iron based superconductors; primarily in their superconducting regime to determine the order parameter magnitude and symmetry. In this review however, we are concentrating on the non-superconducting temperatures and dopings of these compounds. Anomalous signatures have been observed in the normal states of the 122, 11 and 1111 families of the iron based superconductors. 

\subsection{122 Compounds}

\subsubsection{$\rm{Ba(Fe_{1-x}Co_x)_2As_2}$}

\begin{figure*}[tpbh]
		\includegraphics[scale=0.5]{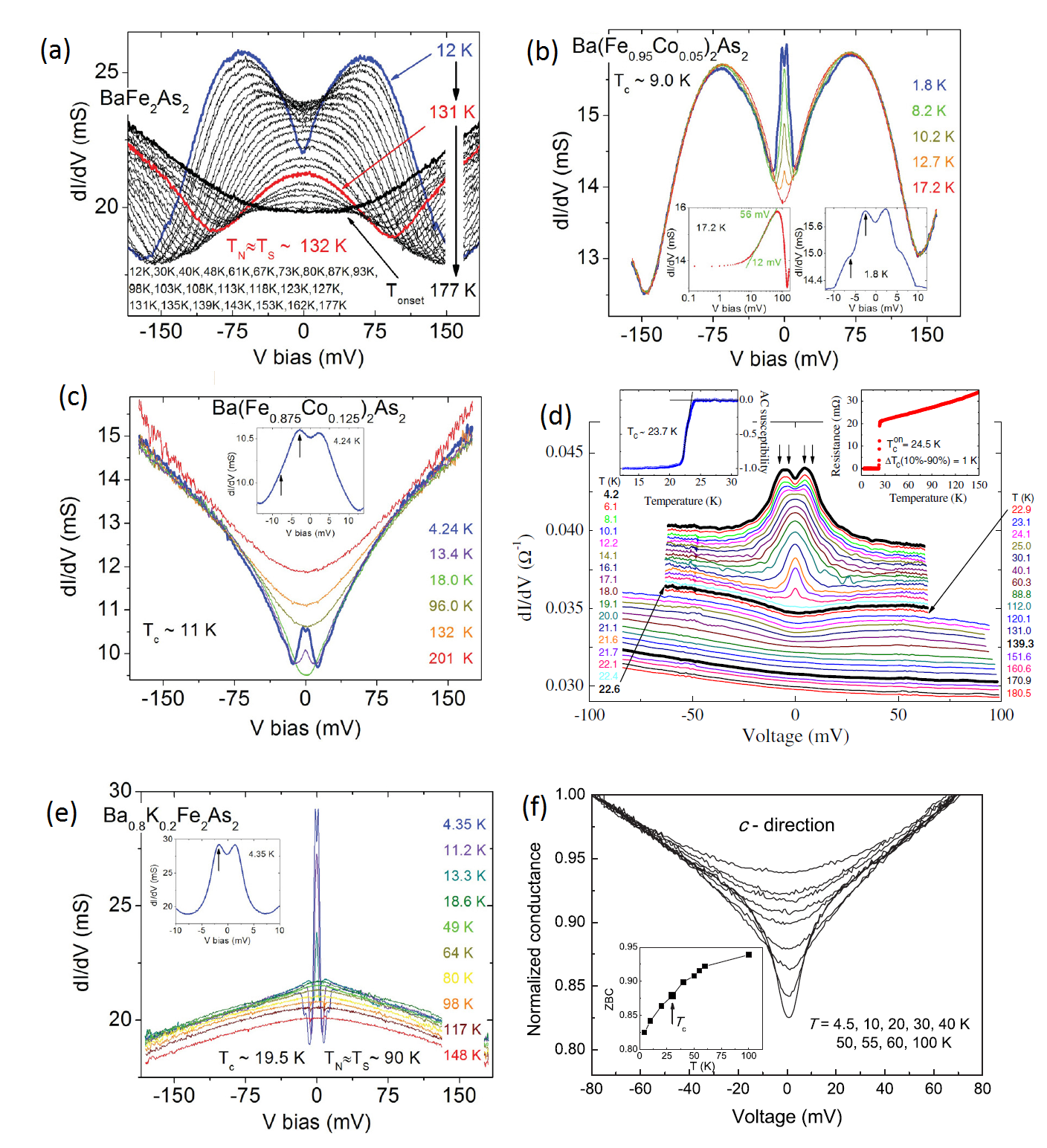}
	\caption{(a) From \cite{Arham}. Conductance spectra for $\rm{BaFe_2As_2}$. Conductance enhancement with peaks at $\sim$ 65 mV superimposed on a parabolic background was observed at low temperatures. The peaks moved in as the temperature was increased and the enhancement survived well above $T_S$ (red curve). (b) From \cite{Arham}. $\rm{Ba(Fe_{0.95}Co_{0.05})_2As_2}$ displayed a coexistence of magnetism and superconductivity. At low temperatures, clear Andreev peaks were observed [right inset (b); the arrows are pointing out the Andreev peaks]. A conductance enhancement with peaks at $\sim$ 65 mV coexisted with the Andreev spectra and evolved with temperature as it did for $\rm{BaFe_2As_2}$. This enhancement increased logarithmically near zero bias [left inset (b)]. (c) From \cite{Arham}. The overdoped compound $\rm{Ba(Fe_{0.875}Co_{0.125})_2As_2}$ showed Andreev spectra below $T_c$. It did not have conductance peaks at higher bias values like the Co underdoped compounds. (d) From \cite{Tortella}. Below $T_c$ the overdoped compound $\rm{Ba(Fe_{0.9}Co_{0.1})_2As_2}$ showed Andreev reflection. Above $T_c$ the spectra was slightly V-shaped and had a dip at zero bias. It flattened out with further increase in temperature. (e) From \cite{Arham}. The hole underdoped $\rm{Ba_{0.8}K_{0.2}Fe_2As_2}$ has a coexistence of superconductivity and magnetism. It showed Andreev spectra below $T_c$ and no higher bias conductance enhancement. This was in contrast to the data obtained from electron underdoped $\rm{Ba(Fe_{1-x}Co_x)_2As_2}$ (Figure 2b). (f) From \cite{Szabo}. Instead of Andreev reflection, the hole overdoped $\rm{Ba_{0.55}K_{0.45}Fe_2As_2}$ sometimes showed a dip around zero bias below $T_c$ that survived well into the normal regime. The might have been due to surface contamination or phase separation in the compound. The inset shows the zero bias conductance versus temperature. }
	\label{fig:311}
\end{figure*}  

Arham et al. \cite{Arham} reported conductance spectra on $\rm{Ba(Fe_{1-x}Co_x)_2As_2}$, spanning the entire phase diagram. Part of their data is reproduced in Figure 2. For the undoped parent compound (Figure 2a), at the lowest temperature (blue curve), they saw a dip at zero bias and two asymmetric conductance peaks at $\sim$ 65 mV. This double peak feature was superimposed on a parabolic background. As the temperature was increased, the dip at zero bias filled up, the conductance peaks moved inward, and the bias voltage range of the conductance enhancement decreased. No dramatic change occurred as $T_S$ was crossed (red curve). The enhancement eventually disappeared at 177 K, more than 40 K above $T_S$. 

For underdoped $\rm{Ba(Fe_{1-x}Co_x)_2As_2}$ (Figure 2b), where superconductivity coexists with long range magnetic order, they observed Andreev reflection at low voltage biases in the superconducting state. However, just like the parent compound, two conductance peaks occurred at $\sim$ 65 mV. Above the onset temperature of the superconducting transition, Andreev reflection completely disappeared and the high bias conductance evolved just like it did for $\rm{BaFe_2As_2}$. The right inset in Figure 2b shows a zoom in of the Andreev reflection features while the left inset plots the conductance spectra above $T_c$ on a log plot. 

For overdoped $\rm{Ba(Fe_{1-x}Co_x)_2As_2}$ (Figure 2c), Andreev reflection was observed in the superconducting state, but unlike the underdoped compounds, no higher bias conductance peaks were detected. Above $T_c$ (the superconducting transition temperature), a parabolic background remained which flattened with further increase in temperature.  

Tortella et al. \cite{Tortella} reported conductance spectra on slightly overdoped $\rm{Ba(Fe_{1-x}Co_x)_2As_2}$ (Figure 2d). Below $T_c$ they observed Andreev reflection with no extraneous features at high bias. Above $T_c$, they were left with a slightly V-shaped curve that progressively filled up on further increase of temperature. This agreed with what Arham et al. observed for strongly overdoped $\rm{Ba(Fe_{1-x}Co_x)_2As_2}$. The insets in the figure show the resistance and AC susceptibility of the compound with temperature.

It appears that underdoped $\rm{Ba(Fe_{1-x}Co_x)_2As_2}$ has rich features in the normal phase; high bias conductance peaks are observed at low temperature that merge together to form a zero bias enhancement that survives beyond the magnetostructural transition. On the other hand, overdoped $\rm{Ba(Fe_{1-x}Co_x)_2As_2}$ has a V shaped parabolic background in the normal state that flattens with increasing temperature. 

\subsubsection{$\rm{Ba_{1-x}K_{x}Fe_2As_2}$}

Since electron underdoped $\rm{Ba(Fe_{1-x}Co_x)_2As_2}$ has rich features in the normal state, an obvious question is what does the normal state spectra on hole underdoped $\rm{Ba_{1-x}K_{x}Fe_2As_2} $ looks like. Arham et al. \cite{Arham} presented data on a $\rm{Ba_{0.8}K_{0.2}Fe_2As_2}$ (Figure 2e). The sample has a coexistence of magnetism and superconductivity ($T_N$ = $T_S$ $\sim$ 90 K, $T_{c}$ $\sim$ 20 K). Below $T_c$ clear Andreev reflection was observed. Above $T_c$, Andreev reflection disappeared leaving a downward shaping background that did not change with any further increase in temperature. This is remarkably different from the situation in electron underdoped $\rm{Ba(Fe_{1-x}Co_x)_2As_2}$. 

Szabo et al. \cite{Szabo} presented data in the normal state of overdoped $\rm{Ba_{1-x}K_{x}Fe_2As_2}$. Sometimes they observed Andreev reflection spectra below $T_c$ (Figure 3 in \cite{Szabo}) while at other times they did not observe Andreev reflection, and instead obtained dips around zero bias that survived well into the normal regime (Figure 2f). The lack of Andreev reflection hints that this might be due to crystal surface contamination or phase separation in the compound. 

\subsubsection{$\rm{CaFe_2As_2}$ and $\rm{SrFe_2As_2}$}

Arham et al. \cite{Arham} also probed $\rm{CaFe_2As_2}$ and $\rm{SrFe_2As_2}$. The data for $\rm{SrFe_2As_2}$ (Figure 3a) was very similar to that of $\rm{BaFe_2As_2}$, with conductance enhancement around zero bias that set in before the magnetostructural transition. However, for $\rm{CaFe_2As_2}$ (Figure 3b), the conductance enhancement disappeared around 100-110 K, much before the magnetostructural transition (170 K). 

\subsection{11 Compounds}

\subsubsection{$\rm{Fe_{1+y}Te}$}

Arham et al. \cite{Arham} presented data on the parent compound for the iron chalcogenide superconductor. $\rm{Fe_{1.13}Te}$ showed a conductance enhancement that survived above the magnetic and structural transition temperatures (Figure 3c). The conductance enhancement was observed till 75 K ($T_N$ = $T_S$ $\sim$ 59 K). 

To summarize the work of Arham et al. \cite{Arham}, $\rm{BaFe_2As_2}$, $\rm{SrFe_2As_2}$, underdoped $\rm{Ba(Fe_{1-x}Co_x)_2As_2}$ and $\rm{Fe_{1+y}Te}$ exhibited a conductance enhancement that set in above $T_S$, $\rm{CaFe_2As_2}$ only showed the enhancement below $T_S$ while $\rm{Ba_{0.8}K_{0.2}Fe_2As_2}$ did not show any conductance enhancement. Overdoped $\rm{Ba(Fe_{1-x}Co_x)_2As_2}$ does not have a $T_S$ and only showed Andreev spectra below $T_c$. The high bias background for all compounds except for $\rm{Ba_{0.8}K_{0.2}Fe_2As_2}$ was an upward facing parabola.

\subsection{1111 Compounds}

\subsubsection{$\rm{SmFeAsO_{1-x}F_x}$}

Daghero et al. \cite{Daghero2} studied the 1111 compound $\rm{SmFeAsO_{0.8}F_{0.2}}$. Below $T_c$ Andreev reflection was present, and above $T_c$, a broad conductance enhancement at zero bias was observed (Figure 3d). The enhancement reduced in amplitude with increasing temperature and disappeared around 140 K, the Neel temperature of the parent compound $\rm{SmFeAsO}$. 

\subsubsection{$\rm{LaFeAsO_{1-x}F_x}$}

Gonnelli et al. \cite{Gonnelli} studied the 1111 compound $\rm{LaFeAsO_{1-x}F_x}$, with a nominal 10$\%$ F doping. At the lowest temperatures, along with Andreev reflection, they observed conductance peaks at higher bias voltages, on the order of 50 mV or more (Figures 3e, f). These peaks reduced in size as the temperature was increased and eventually disappeared close to 140 K, the Neel temperature of the parent compound $\rm{LaFeAsO}$. 

\begin{figure*}[tpbh]
		\includegraphics[scale=0.5]{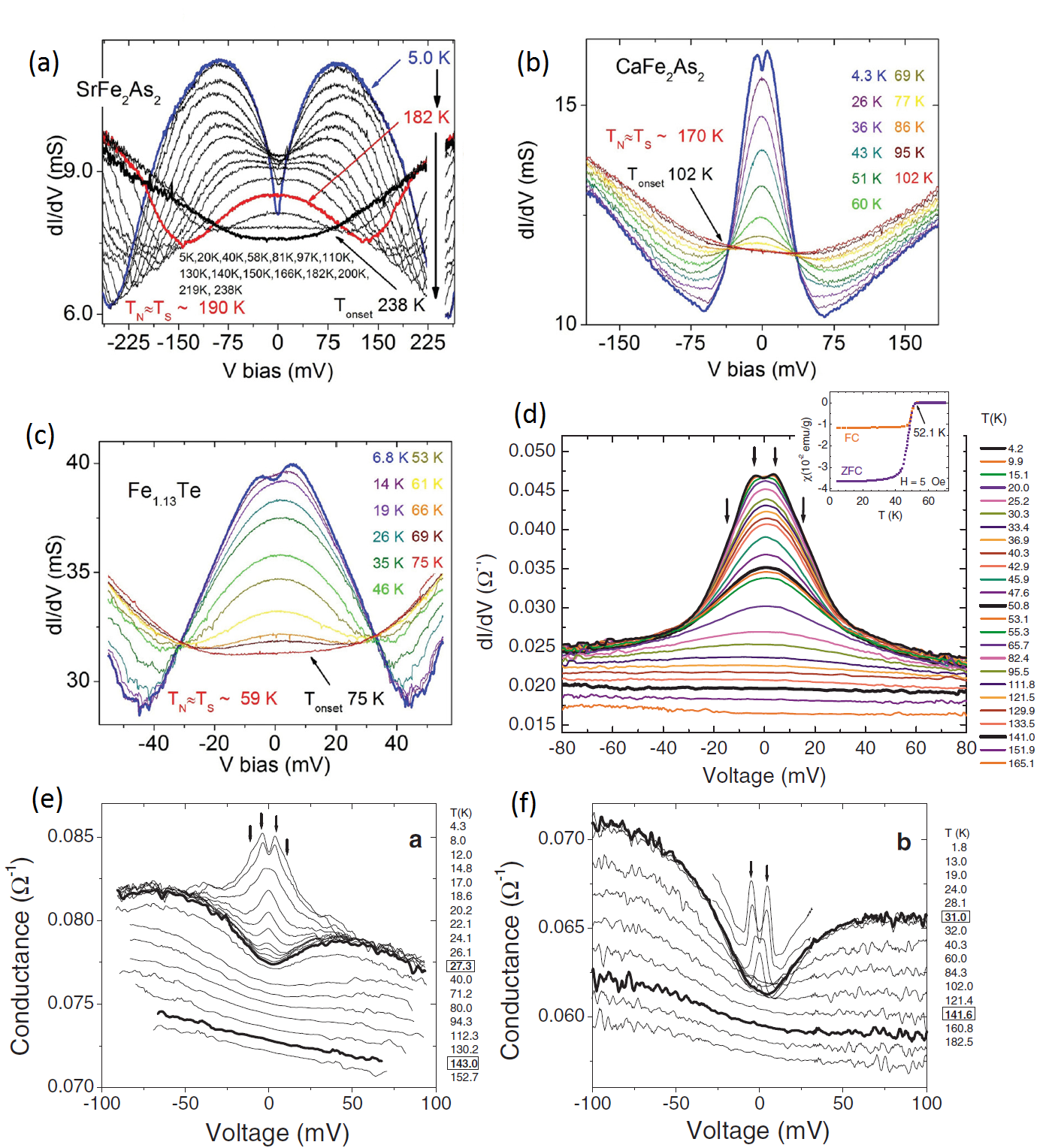}
	\caption{(a) From \cite{Arham}. Conductance spectra for $\rm{SrFe_2As_2}$. The conductance enhancement lasted above $T_S$ and the spectra was similar
to that of $\rm{BaFe_2As_2}$ (Figure 2a). (b) From \cite{Arham}. Conductance spectra for $\rm{CaFe_2As_2}$. In this case the conductance enhancement disappeared before $T_S$. (c) From \cite{Arham}. Conductance spectra for $\rm{Fe_{1.13}Te}$ showed an enhancement that lasted above $T_S$. (d) From \cite{Daghero2}. At the lowest temperature Andreev reflection was observed for $\rm{SmFeAsO_{0.8}F_{0.2}}$, the arrows are pointing out the Andreev peaks. Above $T_c$ ($\sim$ 51 K), a broad conductance enhancement at zero bias was observed. The enhancement reduced in amplitude with increasing temperature and disappeared around 140 K, the Neel temperature of the parent compound $\rm{SmFeAsO}$. (e, f) From \cite{Gonnelli}. At the lowest temperatures, along with the Andreev spectra for $\rm{LaFeAsO_{1-x}F_x}$, (with a nominal 10$\%$ F doping), conductance peaks at higher bias voltages, on the order of 50 mV or more were also observed. These peaks reduced in size as the temperature was increased and eventually disappeared close to 140 K, the Neel temperature of the parent compound $\rm{LaFeAsO}$.}
	\label{fig:311}
\end{figure*}  

\subsection{Discussion of Experimental Results}

\begin{figure*}[tpbh]
		\includegraphics[scale=0.5]{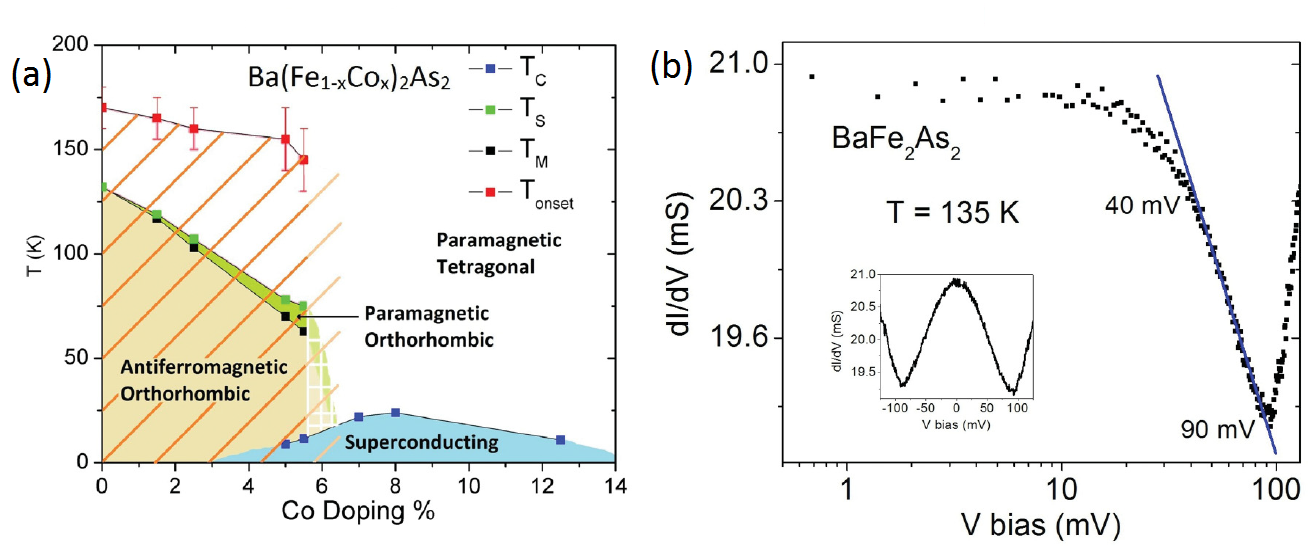}
	\caption{(a) From \cite{Arham}. Phase diagram for $\rm{Ba(Fe_{1-x}Co_x)_2As_2}$ marking a new line on the underdoped side showing the temperature below which the conductance enhancement was observed by PCS. (b) From \cite{Arham}. Conductance above $T_S$ for $\rm{BaFe_2As_2}$ followed a log dependence from $\sim$ 40 mV to $\sim$ 90 mV.}
	\label{fig:311}
\end{figure*}  

At present, we lack a theoretical model to explain the conductance enhancement observed in the normal state of all these compounds. Daghero et al. \cite{Daghero2} speculated that the zero bias conductance enhancement observed in $\rm{SmFeAsO_{1-x}F_x}$ might have a magnetic origin since it disappeared around the Neel temperature of the parent compound. Gonnelli et al. \cite{Gonnelli} speculated that the features they observe in the normal state of $\rm{LaFeAsO_{1-x}F_x}$ might be tied to short range antiferromagnetic fluctuations, as it also disappeared close to the Neel temperature of the parent compound. Szabo et al. \cite{Szabo} raised the possibility of phase separation in these compounds giving rise to the anomalous spectra.

Arham et al. \cite{Arham} noticed a correlation between the presence of conductance enhancement around zero bias and in-plane resistive anisotropy in the compounds. For detwinned underdoped $\rm{AEFe_2As_2}$ it has been shown that below $T_S$ a resistive anisotropy exists \cite{Fisher,Tanatar,Blomberg}. Above $T_S$ there is notable anisotropy for AE = Ba, negligible anisotropy for AE = Sr, and no anisotropy for AE = Ca (Fig. 5 in \cite{Blomberg}). Detwinned $\rm{Fe_{1+y}Te}$ also shows a resistive anisotropy above the structural transition \cite{Jiang}. The anisotropy above $T_S$ is sensitive to the uniaxial force required to detwin the samples. Detwinned underdoped $\rm{Ba_{1-x}K_{x}Fe_2As_2}$ does not show any anisotropy at all, either below or above $T_S$ \cite{Ying}. 

The presence or absence of the in-plane resistive anisotropy matches whether or not a conductance enhancement is detected. The correlation of the conductance enhancement with the resistance anisotropy indicates they are likely caused by the same underlying physics. Arham et al. \cite{Arham} constructed a revised phase diagram for $\rm{Ba(Fe_{1-x}Co_x)_2As_2}$ marking a new line on the underdoped side showing the temperature below which the conductance enhancement was observed (Figure 4a). 

Theoretical work by Lee et al. \cite{WCLee} showed that orbital fluctuations above $T_S$ were expected to provide extra contributions to the single-particle density of states (DOS) at zero energy. The DOS followed a log dependence as the energy was increased. Arham et al. compared their data with this prediction and found that the conductance enhancement for $\rm{BaFe_2As_2}$ above $T_S$ followed a log dependence from $\sim$ 40 mV to $\sim$ 90 mV (Figure 4b). They obtained similar fits above $T_S$ for $\rm{SrFe_2As_2}$ and $\rm{Fe_{1.13}Te}$. Furthermore, the absence of similar effects in the data on $\rm{Ba_{0.8}K_{0.2}Fe_2As_2}$ was consistent with the prediction that crystals that did not show the resistance anisotropy would also not exhibit the excess conductance due to orbital fluctuations. Their data therefore strongly indicated that the enhancement in conductance observed was a consequence of orbital fluctuations.

It should be kept in mind that the conductance ($dI/dV$) measured by point-contact spectroscopy does not directly correspond to the density of states. PCS data is a convolution of the Fermi velocity and the energy-dependent density of states along with any scattering processes that might be present. For normal metals, the Fermi velocity and the density of states are inversely related and cancel each other out \cite{Harrison}. There is a lack of theoretical models for interpreting PCS data on correlated metals, where the DOS are energy dependent and do not cancel out with the Fermi velocity when $dI/dV$ is measured. A theory considering both the energy dependence of the electronic DOS and scattering processes would be extremely helpful in obtaining a better understanding of the experimental data.

\begin{figure*}[hpbt]
		\includegraphics[scale=0.5]{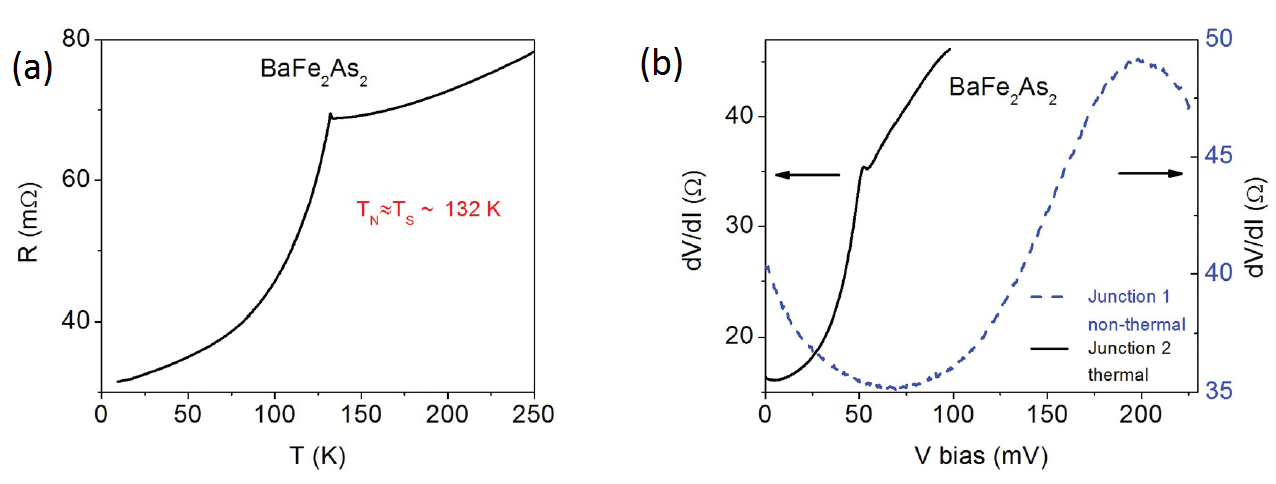}
	\caption{From \cite{Arham}. (a) Resistance vs temperature for $\rm{BaFe_2As_2}$. The bulk resistance always rose with an increase in temperature. A
gradient change occured as the magnetostructural transition was crossed. (b) $dV/dI$ for two junctions on $\rm{BaFe_2As_2}$. Junction 2 was in the thermal limit and followed the functional form of the bulk resistivity (black solid curve, taken at 7.6 K). The junction resistance rose with increasing
voltage and there was a kink at $\sim$ 52 mV corresponding to being heated across the magnetostructural transition. Junction 1 behaved very differently
from bulk resistivity (blue dashed curve, taken at 2.0 K). The junction resistance decreased with an increasing voltage from 0 to $\sim$ 70 mV, and
again for voltages larger than $\sim$ 198 mV. A lack of agreement between bulk resistivity and $dV/dI$ indicated that junction 1 was free of heating
effects.}
	\label{fig:311}
\end{figure*} 

With decreasing temperature, all the excess conductance curves in the data by Arham et al. \cite{Arham} developed a dip at zero bias that sharpened as the temperature was lowered further. Similar dips at zero bias and peaks at high voltages were observed by Gonnelli et al \cite{Gonnelli}. This could happen if there are two dominant scattering processes with opposite voltage dependence at work and the crossover between them giving rise to a peak in $dI/dV$. PCS on Kondo systems shows a similar effect where the Kondo scattering and phonon scattering give rise to a peak in $dI/dV$ \cite{Jansen}. An alternate explanation is that this may be due to the formation of the spin-density-wave (SDW) gap. Previous work has shown PCS to be sensitive to such gapping \cite{Meekes,Escudero}. For $\rm{BaFe_2As_2}$ the conductance peak to peak distance lies between 110 and 140 mV. This agrees well with the SDW gap size (100-125 mV) reported by Raman spectroscopy \cite{Chauviere}, optical conductivity \cite{Nakajima}, and ARPES \cite{Richard} (angle resolved photoemission spectroscopy). However, Gonnelli et al. \cite{Gonnelli} obtained this feature from $\rm{LaFeAsO_{1-x}F_x}$ which does not have a spin density wave present. In addition, the conductance rises logarithmically along the dip (Figure 1b, left inset). These two observations lend credence to the scattering scenario. 

Evidence for normal-state nematicity from detwinned samples is complicated by the symmetry-breaking pressure applied to detwin the crystal. Apart from the resistive anisotropy already discussed, ARPES \cite{Yi} detects orbital ordering, and optical conductivity \cite{Dusza} detects an in-plane anisotropy in the normal state of $\rm{BaFe_2As_2}$. On twinned samples, inelastic neutron scattering reveals high-energy ($>$100 meV) spin excitations above $T_S$ \cite{Harriger}, although that these are truly indicative of nematicity is unclear \cite{Kotliar}. Torque magnetometry on $\rm{BaFe_2(As_{1-x}P_x)_2}$ detects a $C_4$ symmetry breaking in the normal state, across the phase diagram \cite{Matsuda}. Strong anisotropy observed by scanning tunneling microscopy on FeSe, that lacks long-range magnetic order, has been explained using orbital ordering \cite{Song}.

There is an interesting trend between the order of the magnetostructural phase transition and the presence of the conductance enhancement in the normal state. The transition is second order for those compounds that show the enhancement in the normal state ($\rm{BaFe_2As_2}$ \cite{Rotter}, Co doped $\rm{BaFe_2As_2}$ \cite{NiNi2}) while it is first order for those compounds that do not ($\rm{CaFe_2As_2}$ \cite{NiNi} and K doped $\rm{BaFe_2As_2}$ \cite{Avci}).

\section{Non-ideal point contact junctions}

To obtain spectroscopic information from PCS, it is imperative that the junctions are devoid of joule heating effects and any artifacts that may occur due to the junction design. Heating effects will wash out any spectroscopic information while a faulty junction design will display features not representative of the bulk crystal. For point contact on superconductors, the usual test is to check whether the spectra can be explained by the BTK model \cite{Blonder}. Baltz et al. \cite{Baltz} have covered the various unwanted features that may show up in PCS spectra on superconductors, and compared them with ideal PCS spectra.

The case of point contact spectra in the normal state of iron based superconductors is more complicated since there is no BTK like theory to differentiate good point contacts from the bad ones. However, there are other tell-tale signs that may be used to determine the quality of the junction. 

An important characteristic of point contacts in the thermal limit is that $dV/dI$ at low temperature tracks the bulk resistivity. $dV/dI$ is the inverse of the conductance of the point contact junction. Arham et al. (Figure 5) showed that if their point contact junctions were large enough, they did end up in the thermal limit; the spectra followed bulk resistivity and looked completely different from the junctions in the diffusive limit. Figure 5a shows the bulk resistivity of $\rm{BaFe_2As_2}$ while Figure 5b shows $dV/dI$ curves for two junctions, one in the thermal limit, and one in the non-thermal limit. The junction in the thermal limit reproduces the bulk resistivity curve while the non-thermal junction looks completely different. 

The parent iron based compounds undergo a magnetostructural transition that shows up as a sharp gradient change in the bulk resistivity. The thermal limit junction in Figure 5b reproduced a corresponding kink in the conductance spectra at $\sim$ 52 mV while the non-thermal junction did not. This means that when the thermal limit junction was biased at $\sim$ 52 mV, its local temperature had risen to $\sim$ 132 K, the $T_S$ of $\rm{BaFe_2As_2}$.      

Another way to check the reproducibility of the spectra is to form point contact junctions using various methods. Apart from the needle-anvil and soft PCS techniques already described, point contact junctions may be formed via nanolithography or break-junction technique \cite{Naidyuk}. In nanolithography, a counter-electrode smaller than the electron mean free path is patterned onto the compound, while in the break-junction technique, the compound is broken in situ, and then the two pieces are brought together to form a junction. If different methods show similar spectra, it may be concluded that the spectra cannot be an artifact based on the junction design. Arham et al. have obtained similar spectra on $\rm{Fe_{1+y}Te}$ using needle-anvil and soft PCS (Figure 4d in \cite{Arham}). 

Another check for thermal PCS junctions is to compare the $dI/dV$ (V, low T) curve with the zero bias conductance curve, $dI/dV$ (0 mV, T). The local temperature in a thermal junction is related to the bias voltage by $T^2_{max} = T^2_{bath} + V^2/4L$ where L is the Lorentz number of the compound. In such a scenario, there is a substantial overlap between ZBC and $dI/dV$. Arham et al. compared the two quantities for thermal and non-thermal junctions using the appropriate values for the Lorentz number. As expected, for thermal junctions ZBC and $dI/dV$ overlapped while for non-thermal junctions they did not (Figure 6 in \cite{Arham}). 

The parent Fe122 pnictides, Fe11 chalcogenides and $\rm{LaFeAsO_{1-x}F_x}$ have very different resistivity curves. Despite this, at low temperatures in their normal states, they show similar conductance spectra. This provides further evidence that the work of Gonnelli et al. \cite{Gonnelli} and Arham et al. \cite{Arham} is not in the thermal regime and that the same scattering mechanisms are at work in these compounds. Figure 6 \cite{Arham} shows bulk resistance and zero bias conductance curves for $\rm{BaFe_2As_2}$, $\rm{CaFe_2As_2}$ and $\rm{Fe_{1+y}Te}$. The insets show $dI/dV$ at different temperatures. Despite having remarkably different temperature dependence for bulk resistivity, the zero bias conductance and $dI/dV$ for the three compounds follow a similar trend. 

\begin{figure}[hpbt]
		\includegraphics[scale=0.7]{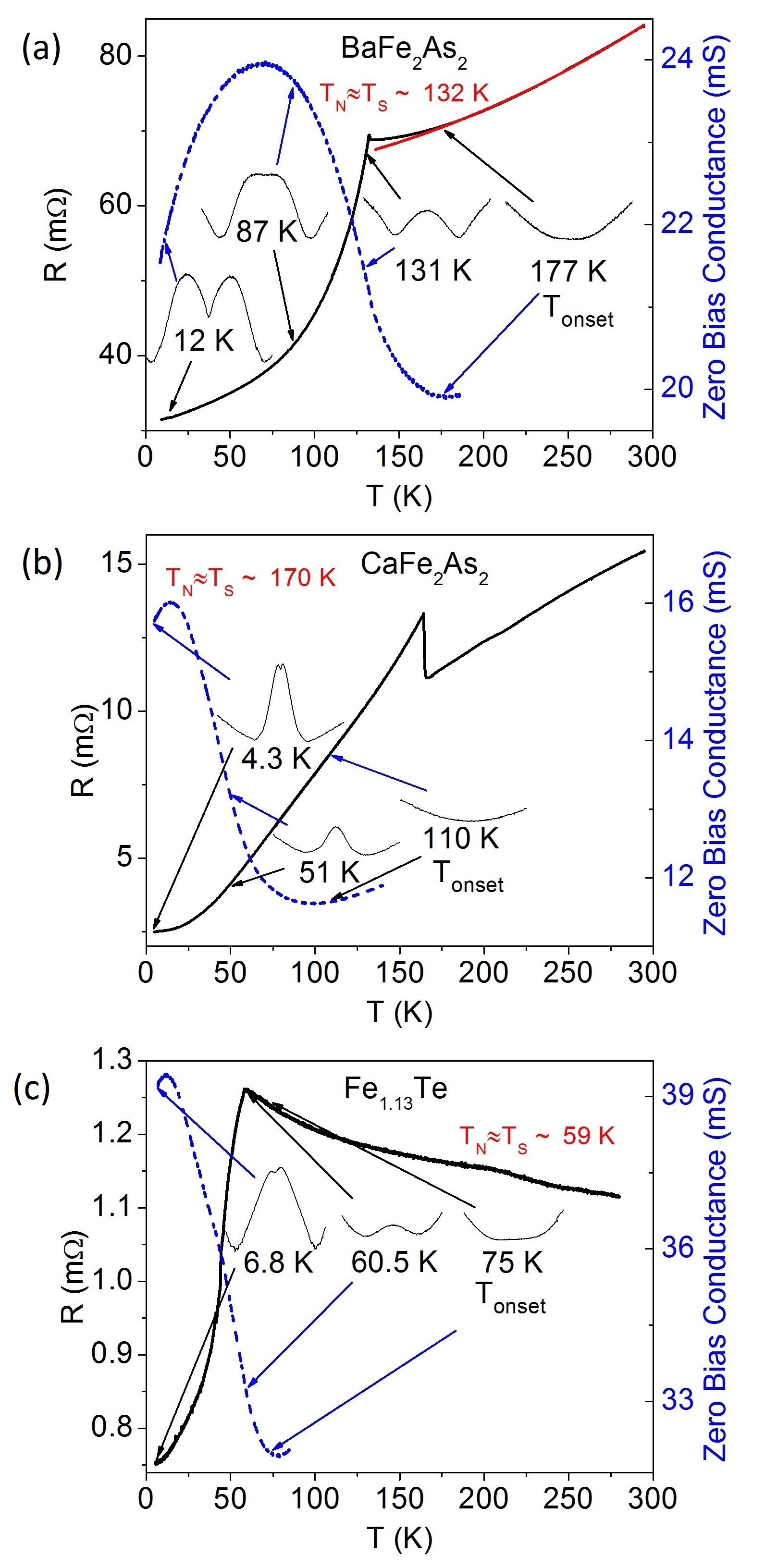}
	\caption{From \cite{Arham}. The zero bias conductance (blue dashed curves) and the resistance vs. temperature, (black solid curves) for $\rm{BaFe_2As_2}$, $\rm{CaFe_2As_2}$ and $\rm{Fe_{1+y}Te}$. For $\rm{CaFe_2As_2}$ the conductance enhancement disappears before $T_S$ while for $\rm{BaFe_2As_2}$ and $\rm{Fe_{1+y}Te}$ it lasts into the normal state. The insets correlate the spectra obtained at different temperatures to the ZBC curve. The red curve in (a) is a fit to $\rho=\rho_0+AT^2$. Despite having remarkably different bulk resistivity properties, the ZBC and $dI/dV$ for the three compounds follow a similar trend.}
	\label{fig:311}
\end{figure}

\section{Outstanding Challenges and Conclusions}

To conclude, point contact spectroscopy in the normal state of certain iron based compounds shows some unique features. $\rm{BaFe_2As_2}$, underdoped $\rm{Ba(Fe_{1-x}Co_x)_2As_2}$, $\rm{SrFe_2As_2}$ and $\rm{Fe_{1+y}Te}$ show a conductance enhancement around zero bias that sets in above the magnetostructural transition. This conductance enhancement might arise due to orbital fluctuations that increase the single particle DOS at zero energy. $\rm{CaFe_2As_2}$ only shows a conductance enhancement below $T_S$, while underdoped $\rm{Ba_{1-x}K_{x}Fe_2As_2}$ does not show any conductance enhancement above $T_c$. The prevalence of the enhancement matches exactly with the presence of an in-plane resistive anisotropy in the bulk compounds. $\rm{SmFeAsO_{1-x}F_x}$ and $\rm{LaFeAsO_{1-x}F_x}$ show similar features that disappear close to the magnetic transition temperatures of their undoped parent compounds. As the temperature is lowered, in all cases apart from $\rm{SmFeAsO_{1-x}F_x}$, the conductance enhancement around zero bias develops a dip. The reason for this behavior is not clear. 

A major challenge is to come up with a microscopic theory for point contact spectroscopy on correlated materials; something that would incorporate the relationship between PCS and the local density of states along with the energy dependent scattering processes intrinsic to these compounds.  

There also needs to be more experimental work done on the iron based superconductors. It would be interesting to know how the isoelectronic doped $\rm{BaFe_2(As_{1-x}P_x)_2}$ behaves in the normal state. Arham et al. have preliminary data indicating that a conductance enhancement is present in the normal state of $\rm{BaFe_2(As_{1-x}P_x)_2}$ and $\rm{NaFe_{1-x}Co_xAs}$. In addition, PCS needs to be performed on these compounds under applied magnetic field and when the crystals have been detwinned by applied pressure. 

Overall, point contact spectroscopy provides strong indication for electronic nematicity arising from orbital fluctuations in those iron based compounds that have an in-plane resistive anisotropy in the normal state. This matches up with what has been observed by ARPES, optical conductivity and inelastic neutron scattering in these compounds. 

\section*{Acknowledgments}

This work was supported by the Center for Emergent Superconductivity, an Energy Frontier Research Center funded by the US Department of Energy, Office of Science, Office of Basic Energy Sciences under Award No. DE-AC0298CH1088.

\end{document}